%% file: iclr2026_conference.tex
\documentclass[11pt,letterpaper,logo]{mystyle}

\usepackage{natbib}
\usepackage{multirow}
\usepackage{float}
\usepackage{wrapfig}
\usepackage{fontawesome}

\title{RSIBench-Data: Benchmarking Data-Centric Research for Recursive Self-Improvement}
\runningtitle{RSIBench-Data}

\author{Fanqing Meng$^{1,2,*}$, Lingxiao Du$^{1,2,*}$, Qiguang Chen$^{1,*}$, Ziqi Zhao$^{1}$, Haocheng Lu$^{1}$,
Mengkang Hu$^{1,\dagger}$, Michael Qizhe Shieh$^{1,2,\dagger}$\\
{\small $^{1}$Evolvent AI \quad $^{2}$National University of Singapore}\\
{\small $^{*}$Equal contribution. $^{\dagger}$Corresponding authors.}\\
{\fontsize{9pt}{12pt}\selectfont
\href{https://github.com/evolvent-ai/RSIBench-Data}{\faGithub~\,github.com/evolvent-ai/RSIBench-Data}\hspace{14pt}%
\href{http://rsibench.co}{\faGlobe~\,rsibench.co}}}

\newcommand{\benchmark}{\textsc{RSIBench-Data}}

\newcommand{\swebench}{\textsc{SWE-bench}}
\newcommand{\tinker}{\textsc{Tinker}}
\newcommand{\harbor}{\textsc{Harbor}}
\newcommand{\eibtwo}{\textsc{E2B}}
\newcommand{\miniswe}{\textsc{Mini-SWE-Agent}}

\begin{document}

\begin{abstract}
\input{sections/abstract}
\end{abstract}

\maketitle
\thispagestyle{firststyle}

\input{sections/introduction}
\input{sections/related_work}
\input{sections/dataset}
\input{sections/experiment}
\input{sections/results}
\input{sections/discussion}
\input{sections/conclusion}

\bibliography{references}
\bibliographystyle{plainnat}

\appendix
\input{sections/appendix_action_space}
\end{document}

%% file: sections/abstract.tex
Recursive self-improvement requires systems that can turn evidence about model failures into better models. A central part of this process is data-centric post-training research: diagnosing a capability gap, designing and validating a training-data strategy, and learning from checkpoint feedback. Can LLM agents automate this data-centric improvement loop? Existing automated post-training benchmarks entangle research decisions with optimization, serving, evaluation, and system implementation, making it difficult to isolate the agent's research capability. To fill this gap, we introduce \benchmark{}, a controlled benchmark evaluating LLM agents as data-centric researchers while fixing the post-training stack. Specifically, agents are required to iteratively develop and revise training-data strategies for a fixed target model, while training and serving use shared \tinker{}-backed services, official evaluation runs through \harbor{}-orchestrated \eibtwo{} sandboxes, and resource budgets remain fixed across agents. We evaluate four frontier agents across six benchmarks spanning software engineering, terminal use, science question answering, and mathematics. LLM agents already demonstrate some of the core capabilities of data-centric researchers: in 58.33\% of settings, they improve upon the first valid attempt by refining their training-data strategies from feedback. However, they do not improve consistently from feedback: among searches that continue after reaching their best observed score, 78.26\% end with a lower-scoring final attempt, while the rest only recover the same peak. Thus, a strong candidate can appear early or midway through a run even as later feedback-driven revisions fail to improve it. Trajectory analysis identifies four patterns in stronger runs: accurate hypotheses, validation-grounded supervision, behavior-aligned data, and preservation of strong checkpoints. These findings suggest that current agents can make useful data-centric research discoveries but cannot yet translate feedback into improvements consistently. Overall, \benchmark{} provides a measurable and auditable testbed for studying the data-centric research capabilities required for recursive self-improvement. We opensource our code at \url{https://github.com/evolvent-ai/RSIBench-Data}.

%% file: sections/introduction.tex
\section{Introduction}

Recursive self-improvement requires an AI system to repeatedly turn evidence about model failures into effective model improvements. Recently, LLM agents are increasingly evaluated on tasks that require long-horizon interaction with software repositories, terminals, and external tools \citep{swebench2024,merrill2026terminalbenchbenchmarkingagentshard,taubench2024}. Unlike static input--output tasks, these environments expose the agent's intermediate plans, tool calls, observations, recovery attempts, and verifier outcomes \citep{guo2024large,chen2025ai4research,chen2026towards}. The resulting trajectories make failures richly observable, but they remain evidence about model behavior rather than training experience that can directly improve it \citep{barke2026agentrx}.
To address this gap, recent work addresses different parts of this evidence-to-training pipeline \citep{datacomp2023,swegym2024,posttrainbench2026}. SWE-Gym\citep{swegym2024} and R2E-Gym~\citep{r2egym2025} provide executable environments and training resources for software-engineering agents. DataComp~\citep{datacomp2023} and DataComp-LM~\citep{datacomplm2024} study data selection under a controlled learning stack. PostTrainBench~\citep{posttrainbench2026} and Agent$^2$ RL-Bench~\citep{agent2rlbench2026} instead evaluate broader automated post-training systems.

\begin{figure}[t]
    \centering
    \includegraphics[width=0.98\linewidth]{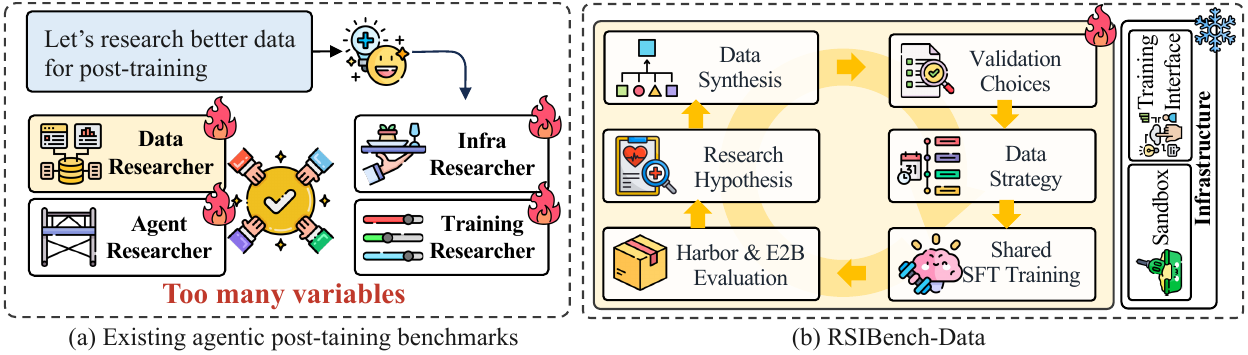}
    \caption{Existing evaluations entangle multiple post-training choices; \benchmark{} isolates the LLM agent's ability to conduct data-centric post-training research.}
    \label{fig:motivation}
\end{figure}

These paradigms address broader questions spanning data selection, environment construction, and end-to-end post-training, but they do not isolate a central research problem for recursive self-improvement: \emph{can LLM agents reliably automate data-centric post-training research?} Specifically, to achieve this goal, the agent must formulate a hypothesis about a model failure, turn it into executable and verifiable training data, and revise the synthesis strategy from controlled training and evaluation feedback \citep{swegym2024,r2egym2025}. We call an LLM agent that carries out this closed-loop process a \emph{data-centric researcher agent}.
However, existing benchmarks entangle research decisions with optimization, serving, evaluation, and system implementation, making it difficult to isolate the agent's research capability (Figure~\ref{fig:motivation}(a)).

To address this gap, we introduce \benchmark{}, which isolates the data-centric research component of recursive self-improvement under a fixed surrounding stack (Figure~\ref{fig:motivation}(b)). Specifically, we fix the base model and bounded training interface, expose training and serving through shared \tinker{} services, and isolate official evaluation behind \harbor{}-orchestrated \eibtwo{} sandboxes under a common budget.
More importantly, \benchmark{} evaluates an iterative training-data research process, not imitation-based dataset generation. In each round, the agent decides what the target model should learn, develops and validates a training-data strategy, trains a checkpoint through shared LoRA SFT, and uses controlled evaluation feedback to choose the next strategy.

Our results reveal a \textbf{discovery--reliability gap}. Across four frontier agents and six benchmarks, agents improve on their first valid attempt in 58.33\% of settings by refining training-data strategies from feedback. This result shows that agents already exhibit some core capabilities of data-centric researchers. However, these gains are task dependent and fragile. No agent consistently dominates: Codex \texttt{gpt-5.6-sol} leads the three non-SWE benchmarks, but the three SWE-style tasks are still won by three different agents. Moreover, agents do not improve consistently from feedback: among searches that continue after reaching their best observed score, 78.26\% finish with a lower-scoring final attempt, while the remainder only recover the same peak. An early or intermediate attempt can therefore remain strongest even after the agent observes additional feedback and revises its strategy. The round-by-round records reveal recurring failures: agents misdiagnose target capabilities, generate misaligned supervision, search past strong checkpoints, or fail to turn new feedback into a stronger training-data strategy.

Overall, our contributions are threefold:
\begin{itemize}[leftmargin=*,topsep=0pt]
    \item \textbf{A new evaluation target: data-centric research for recursive self-improvement.} We formulate post-training data synthesis as an automatable research loop in which an LLM agent forms capability hypotheses, designs and validates training-data strategies, learns from checkpoint evidence, and manages candidate selection under a fixed budget.

    \item \textbf{A controlled benchmark for data-centric researcher agents.} We introduce \benchmark{}, with shared \tinker{} training and serving and official evaluation through \harbor{} and \eibtwo{} sandboxes. These boundaries make experiments auditable and separate infrastructure changes from research progress.

    \item \textbf{Empirical evidence: researcher capabilities emerge, but remain fragile.} Agents improve on the first valid attempt in 58.33\% of settings, yet among searches that continue after reaching a peak, 78.26\% finish with a lower-scoring attempt and the rest only recover that peak. This non-monotonicity shows that agents do not consistently turn feedback into better training-data strategies. Stronger runs feature accurate hypotheses, validation-grounded supervision, behavior-aligned data, and checkpoint preservation.
\end{itemize}

%% file: sections/related_work.tex
\section{Related Work}

\providecommand{\yesmark}{\textcolor{green!55!black}{\ding{51}}}
\providecommand{\nomark}{\textcolor{red!70!black}{\ding{55}}}

% \vspace{-8pt}

\begin{table}[t]
\caption{Protocol-level comparison of benchmarks across data-synthesis research, closed-loop evolution, and controlled infrastructure. A ``\yesmark'' denotes a criterion explicitly required and enforced by the benchmark protocol, rather than merely enabled by available tools.}
\label{tab:positioning}
\centering
\footnotesize
\renewcommand{\arraystretch}{1.12}
\resizebox{\linewidth}{!}{%
\begin{tabular}{@{}lcccccc@{}}
\toprule
& \multicolumn{2}{c}{\textbf{Data-Synthesis Research}} & \multicolumn{2}{c}{\textbf{Closed-loop Evolution}} & \multicolumn{2}{c}{\textbf{Controlled Infrastructure}} \\
\cmidrule(lr){2-3}\cmidrule(lr){4-5}\cmidrule(lr){6-7}
\textbf{Benchmark}
& \shortstack{Executable Experience\\Synthesis}
& \shortstack{Reusable Data-\\Synthesis Policy}
& \shortstack{Capability-gap\\Diagnosis}
& \shortstack{Feedback-driven\\Revision}
& \shortstack{Train/Eval Data\\Isolation}
& \shortstack{Service-isolated\\Train/Serve/Eval} \\
\midrule
DataComp & \nomark & \nomark & \nomark & \nomark & \nomark & \nomark \\
DataComp-LM & \nomark & \nomark & \nomark & \nomark & \nomark & \nomark \\
AgoraBench & \nomark & \nomark & \nomark & \nomark & \nomark & \nomark \\
DCA-Bench & \nomark & \nomark & \nomark & \nomark & \nomark & \nomark \\
\midrule
DataEnvGym & \nomark & \nomark & \yesmark & \yesmark & \nomark & \nomark \\
Curation-Bench & \nomark & \nomark & \nomark & \yesmark & \nomark & \nomark \\
\midrule
PostTrainBench & \nomark & \nomark & \yesmark & \yesmark & \nomark & \nomark \\
Agent$^2$ RL-Bench & \yesmark & \nomark & \nomark & \nomark & \nomark & \nomark \\
\midrule
\textbf{\benchmark{} (ours)} & \yesmark & \yesmark & \yesmark & \yesmark & \yesmark & \yesmark \\
\bottomrule
\end{tabular}
}
\end{table}

\paragraph{Data curation and generation benchmarks.} Early data-centric benchmarks control the surrounding learning stack to study which examples should be selected or curated from a fixed corpus \citep{datacomp2023,datacomplm2024}. To move beyond a provided candidate pool, subsequent work evaluates generated datasets or whether agents can diagnose hidden data defects \citep{agorabench2025,dcabench2024}. DataEnvGym and Curation-Bench are the closest to our closed-loop setting: they introduce student-model or benchmark feedback so that agents can adapt curricula and curation strategies across rounds \citep{dataenvgym2025,curationbench2026}. \benchmark{} extends this direction by evaluating one researcher agent that owns the data-synthesis hypothesis and revision process while the outer training and evaluation loop remains fixed.\vspace{-8pt}

\paragraph{Recursive self-improvement and agent training.} To address the shortage of grounded training experience for tool-using agents, early work~\citep{agentbank2024,agenttrek2024} collects interaction trajectories or provides executable environments for generating such experience \citep{swegym2024,r2egym2025}. Building on these resources, recent systems~\citep{karl2026,socraticswe2026,ict2026} automate experience acquisition, using knowledge exploration, historical execution traces, or self-generated interactions to construct supervision and improve the agent across rounds \citep{seal2026,qevolve2026}. These feedback-driven systems provide mechanisms relevant to recursive self-improvement, but generally evaluate a particular method rather than comparing how different LLM agents perform the same data-centric researcher role under controlled experimental services. \benchmark{} instead isolates the data-centric research component through which a frontier researcher agent improves a separate target model.\vspace{-8pt}

\paragraph{Automated post-training and agentic evaluation.} Automated post-training benchmarks test whether agents can improve other models by granting control over data, optimization, hyperparameters, and implementation \citep{posttrainbench2026,agent2rlbench2026}. Complementary work studies persistent multi-agent evolution, develops fresh contamination-resistant evaluations, and provides long-horizon SWE, terminal, and tool-use targets \citep{coral2026,swerebench2025,swebenchlive2025,swebench2024,merrill2026terminalbenchbenchmarkingagentshard,taubench2024}. Together, these benchmarks measure whether an end-to-end system improves and whether the resulting model succeeds, but their broad scope obscures the contribution of data-centric research decisions.

Existing benchmarks make a fundamental tradeoff: some isolate individual data operations, while others give agents broad control over the post-training stack. \benchmark{} occupies the middle ground: it evaluates whether an LLM agent can act as a data-centric researcher while the surrounding training, serving, and official evaluation stack remains fixed (Table~\ref{tab:positioning}). Rather than scoring only the resulting checkpoint, it records the hypotheses, training-data strategies, feedback-driven revisions, and checkpoint choices that produce it. Shared \tinker{} training and serving and fixed \harbor{}--\eibtwo{} evaluation make this research process auditable.

%% file: sections/dataset.tex
\section{\benchmark{}: Evaluation Task and Protocol}
\label{sec:protocol}

Conventional data-generation evaluations ask whether a system can produce a useful dataset or checkpoint. \benchmark{} asks a research question central to recursive self-improvement: can an LLM agent form a useful hypothesis about a fixed model, design a targeted training-data strategy, interpret checkpoint evidence, and preserve its best finding under a finite budget?

\subsection{Problem Definition}
\label{sec:problem}
We evaluate data-centric research capability through the final checkpoint and the experimental process that produces it. We study a bounded, cross-model setting in which a frontier researcher agent iteratively improves a separate target model through data-centric post-training. Official checkpoint performance remains the primary outcome; the recorded process supports attribution and diagnosis rather than a separate subjective score. \emph{Training experience} includes examples, executable tasks and states, tool-use trajectories, and verifier-grounded supervision.
Formally, given a fixed base model $M_0$, benchmark evidence $S$, and resource budget $\mathcal{C}$, a researcher policy $\pi$ observes $S$ and its attempt history $H_{<t}$, then proposes training data $D_t$ and a whitelisted training config $c_t$:
\begin{equation}
    (D_t,c_t)=\pi(S,H_{<t}), \qquad
    M_t=\operatorname{Train}(M_0,D_t;c_t).
\end{equation}
Each $M_t$ is evaluated by a fixed evaluation service, producing permitted selection feedback $h_t=\operatorname{Eval}_{\mathrm{sel}}(M_t)$, such as scores, trajectories, verifier outcomes, and execution diagnostics. Proposals and feedback form $H_t$, which the agent uses to revise or stop within budget $\mathcal{C}$.

\begin{figure}[t]
    \centering
    \includegraphics[width=\linewidth]{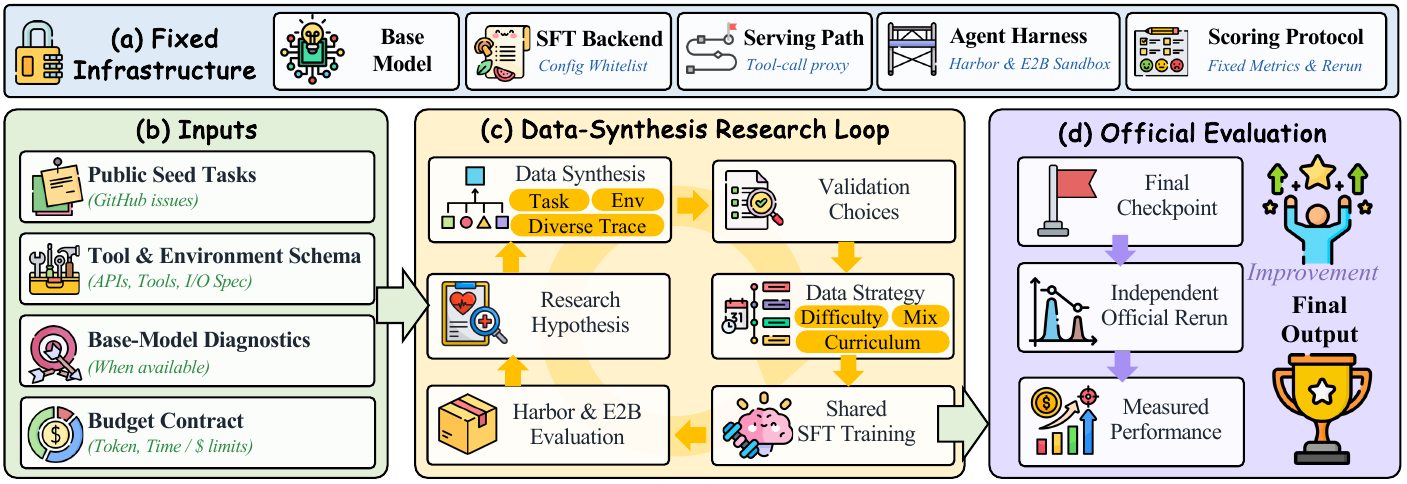}
    \caption{Overview of \benchmark{}, which evaluates how an LLM researcher agent evolves a training-data research policy under shared training, serving, evaluation, and budget controls.}
    \label{fig:framework}
\end{figure}

After $T$ realized attempts, the agent selects a checkpoint $M_{t^\star}$ using only $H_T$, before observing official outcomes. The evaluator then assesses it in fresh task environments without further training. We report its official performance and, secondarily, improvement over the fixed base model:
\begin{equation}
    s_{\mathrm{off}}(\pi)=\operatorname{Eval}_{\mathrm{off}}(M_{t^\star}),
    \qquad
    \Delta_{\mathrm{off}}(\pi)=s_{\mathrm{off}}(\pi)-\operatorname{Eval}_{\mathrm{off}}(M_0).
\end{equation}
Two separations support attribution: agents modify training experience through the bounded interface but not the surrounding stack, and use selection feedback, not official outcomes, for iteration and checkpoint selection. These separations are instantiated through fixed infrastructure, benchmark inputs, a closed-loop data-synthesis research process, and a separate official evaluation (Figure~\ref{fig:framework}).

\subsection{Fixed Infrastructure}
\label{sec:fixed_infrastructure}

\benchmark{} fixes the infrastructure that turns a training-data strategy into an evaluated checkpoint, so that a researcher agent works inside a fixed boundary rather than over the whole post-training stack. Inside the boundary, the agent decides what training experience to construct and, through a bounded configuration interface, how to expose it to training. Outside it, the base model, optimizer, serving path, evaluation sandbox, verifier, and scoring rule are fixed and shared by every agent: the agent submits training artifacts rather than training code, and may not replace any component of the surrounding stack.

Every attempt therefore follows the same path from submitted data to evaluated checkpoint. In each round, the agent submits message-format supervision together with a whitelisted training configuration. The runner validates these artifacts and invokes the same \tinker{}-backed LoRA SFT backend from the fixed base model, and the resulting checkpoint is served through the same sampler interface that the evaluator uses to query it. Execution is then delegated to \harbor{} in \eibtwo{} sandbox environments through a fixed service boundary that owns the benchmark environment, runner, verifier, and scoring rule, and returns controlled evidence to the agent. Optimization, serving, and evaluation are thus identical across agents, while the submitted data distribution and the configuration within the whitelist remain free to vary.

Because data submission, training, serving, and evaluation all pass through this boundary, each attempt leaves a structured record that links a training-data strategy to its checkpoint and evaluation behavior. This makes the loop auditable: we can analyze why a run improves or regresses, and attribute performance differences to the agent's training-data research policy rather than to changes in the post-training stack.

\subsection{Agent Inputs and Data-Use Constraints}
\label{sec:integrity}

Each agent receives the target benchmark description, success criteria, task-matched public seed repositories or seed examples, tool and environment schemas, available base-model diagnostics, and a resource budget. These inputs form $S$ without providing a ready-made training corpus (Figure~\ref{fig:framework}(b)). Seed repositories provide realistic task structure and a reference for the target distribution: without such grounding, synthesis can collapse to simple rule-based examples that are detached from real tasks and circumvent the intended research challenge.

Not all information available to the agent may be included in $D_t$. Each benchmark specifies which seeds and public sources may support data construction; evaluation-only tasks, labels, trajectories, and other protected benchmark materials may not be used as supervision. This separation allows the agent to diagnose failures without converting evaluation information into training data.

\subsection{Closed-Loop Data-Synthesis Research}

\benchmark{} fixes only the outer sequence and service boundaries of the budgeted loop: the agent synthesizes and submits data, the benchmark trains a checkpoint through the shared SFT service, and the fixed sandboxed evaluator returns permitted signals (Figure~\ref{fig:framework}(c)). Filtering, verification, curriculum design, data mixing, failure attribution, and the decision to continue searching are not prescribed benchmark stages; they are choices available to the researcher agent's data-synthesis policy. The budget constrains total resources, while the agent decides how training and evaluation feedback should change the next training-data strategy. Appendix~\ref{app:action_space} describes representative strategies rather than mandatory phases.\vspace{-8pt}

\paragraph{Data synthesis.}
Within each round, the agent may choose data sources, task and environment construction, trajectory representation, filtering rules, verification methods, difficulty curricula, data mixtures, and training exposure. The benchmark neither requires a fixed preliminary diagnosis nor prescribes an internal order for these operations; it requires only that the submitted artifacts satisfy the data contract and can be consumed by the shared SFT interface. After training and evaluation, the agent decides whether to continue and how to evolve its next training-data strategy.\vspace{-8pt}

\paragraph{Shared SFT training.}
The benchmark validates the submitted artifacts and configuration, then trains a LoRA checkpoint from $M_0$ through the shared SFT backend. All agents use the same training implementation and the same bounded configuration interface. This converts each proposed training distribution into a checkpoint without allowing the agent to replace the optimization backend or otherwise redesign the learning stack.\vspace{-8pt}

\paragraph{Harbor and E2B evaluation.}
The checkpoint is queried through the fixed serving path and evaluated by \harbor{} in \eibtwo{} sandbox environments with the official verifier and scoring protocol. The service returns the permitted selection score together with task trajectories, verifier outcomes, execution and infrastructure errors, token usage, training and sampling cost, and elapsed time. Because evaluation is fixed, these signals provide comparable evidence about the submitted training-data strategy.\vspace{-8pt}

\subsection{Final Checkpoint and Official Evaluation}
\label{sec:scoring}

After completing its attempts, the agent selects one checkpoint from those it has produced. We evaluate the selected checkpoint on the target benchmark's test set using the fixed evaluation protocol and report the resulting performance as the agent's final score.

%% file: sections/experiment.tex
\section{Experimental Setup}
\label{sec:experiments}

To isolate data-centric research capability, we instantiate a controlled
$4\times6$ matrix: the researcher agent and target benchmark vary, while the
target model, bounded training interface, evaluation services, and per-run
budget remain fixed.

\subsection{Researcher agents}
We evaluate four frontier LLM researcher agents in the main benchmark matrix: Claude Code~\citep{claudecode2025} with \texttt{Opus-4.8} and \texttt{Sonnet-5}, and Codex~\citep{openaicodex2025} with \texttt{gpt-5.6-sol} and \texttt{gpt-5.6-terra}. The Claude Code agents run at high reasoning effort, while the Codex agents run at max reasoning effort; all agents receive the same task interface, data-use constraints, and budget. Across all experiments, Claude Opus 4.8 is fixed as the external rollout model for generating reasoning traces, tool-use sequences, or full trajectories. Agents submit training data $D_t$ together with a whitelisted configuration $c_t$; our comparisons therefore evaluate data-centric research through the shared bounded interface of Section~\ref{sec:protocol}, rather than data alone under a fully frozen training recipe. We separately include a high-versus-max reasoning-effort comparison for one Claude Code \texttt{Sonnet-5} setting in Section~\ref{sec:research_mechanisms}.

\subsection{Shared training pipeline}
Each attempt trains a LoRA-adapted checkpoint of \texttt{Qwen/Qwen3.5-35B-A3B-Base}~\citep{qwen3technical2025} through the shared \tinker{} SFT backend~\citep{tinker2025}\footnote{\url{https://github.com/thinking-machines-lab/tinker}.}. A \harbor{}-orchestrated evaluation service executes each checkpoint in \eibtwo{} sandboxes and returns the controlled evidence used for subsequent research decisions and checkpoint choice. The selected checkpoint is then evaluated without additional training through the same fixed service boundary\footnote{\url{https://github.com/e2b-dev}.}. Each main run has a nominal 16-hour wall-clock budget and a \$500 \tinker{} budget.

\subsection{Benchmarks and evaluation}
We select six benchmarks spanning distinct training-experience regimes. The three \swebench{} variants---Verified~\citep{swebench2024}, Multilingual~\citep{zan2025multiswebenchmultilingualbenchmarkissue}, and Pro~\citep{deng2025swebenchproaiagents}---require repository-grounded supervision for code modification and task completion. Terminal-Bench~2.0~\citep{merrill2026terminalbenchbenchmarkingagentshard} emphasizes long-horizon tool use and verification, whereas GPQA Diamond~\citep{rein2023gpqagraduatelevelgoogleproofqa} and AIME~2026\footnote{\url{https://huggingface.co/datasets/MathArena/aime_2026}.} emphasize verifiable reasoning and difficulty control.

For each benchmark, all agents share the same evaluation subset, runner, verifier, and parameters. The \swebench{} variants use \miniswe{}~\citep{yang2024sweagent}\footnote{\url{https://github.com/SWE-agent/mini-swe-agent}.}, while the remaining benchmarks use \textsc{Terminus-2}~\citep{merrill2026terminalbenchbenchmarkingagentshard}. The three \swebench{} variants and GPQA~Diamond use fixed 100-instance subsets; Terminal-Bench~2.0 and AIME~2026 use their full sets, with four decoding runs per AIME problem. Evaluation hyperparameters, including temperature, follow each benchmark's official settings. Appendix~\ref{app:evaluation_suite} reports the full evaluation suite and the public seed repositories available to the researcher agents.

The primary outcome is the selected checkpoint's official benchmark-native score. Secondary analyses track improvement over the fixed base model, within-run selection trajectories, the first and best candidates, training and sampling cost, and wall-clock time.

%% file: sections/results.tex
\section{Results and Analysis}
\label{sec:results}

\subsection{Main Results}
\label{sec:overall_results}

\benchmark{} exposes two sides of autonomous data-centric research. First, iteration can produce genuine discoveries: later candidates improve over the first valid candidate in 14 of 24 settings. Second, agents do not consistently convert subsequent feedback into further gains: among the 23 settings that continue after reaching an observed peak, 18 finish with a final attempted candidate below their historical best and 5 only return to the same peak. Historical-best checkpoint selection can preserve an earlier discovery despite these later regressions, but the trajectories still show that feedback-driven revisions often fail to refresh the candidate frontier. We therefore analyze the benchmark in three stages: task-dependent official performance, the consistency and efficiency of feedback-driven improvement, and the research decisions associated with stronger trajectories.\vspace{-8pt}

\paragraph{Performance analysis.}
Official scores and resource profiles for every agent--benchmark setting are summarized in Table~\ref{tab:official_results}. Each score comes from a pre-specified representative run. Performance remains weak on several executable tasks: the best agent reaches only 9.00\% on \swebench{} Pro, 22.00\% on \swebench{} Multilingual, and 20.22\% on Terminal-Bench~2.0. These row-wise ceilings indicate a shared weakness across current researcher agents. Some trained checkpoints also fail to improve over the base model, especially on GPQA~Diamond and \swebench{} Multilingual, showing that training-data strategies can be neutral or harmful as well as helpful. At the same time, no researcher agent dominates the suite. Claude Code \texttt{Opus-4.8} leads on \swebench{} Verified, Claude Code \texttt{Sonnet-5} on \swebench{} Multilingual, and Codex \texttt{gpt-5.6-sol} on \swebench{} Pro, GPQA~Diamond, AIME~2026, and Terminal-Bench~2.0. The three SWE-style tasks are still won by three different agents, showing that broad task-family similarity does not induce a stable ranking. Among the four trained agents, the best-to-worst spread further confirms the strength of this interaction: it is 8.00 percentage points on \swebench{} Pro, 13.00 on GPQA~Diamond and \swebench{} Verified, 14.60 on Terminal-Bench~2.0, 17.00 on \swebench{} Multilingual, and 20.00 on AIME~2026. Because every run uses the same target model and bounded interface, these gaps reflect the interaction between researcher-agent policy and benchmark structure, not variation in the target model or evaluation stack.\vspace{-8pt}

\begin{table}[H]
\caption{Official performance and resource use for the main Agent--Benchmark matrix. The base-model row reports the unadapted \texttt{Qwen/Qwen3.5-35B-A3B-Base} score under the same evaluation split and sampling protocol. Time and \tinker{} cost report the resources used by each setting. Bold marks the best official score within each benchmark.}
\label{tab:official_results}
\centering
\scriptsize
\resizebox{0.8\linewidth}{!}{%
\begin{tabular}{llrrr}
\toprule
Benchmark & Agent & Official & Time (h) & \tinker{} cost (\$) \\
\midrule
\multirow{5}{*}{\swebench{} Verified} & Base model & 12.00\% & -- & -- \\
 & Claude Code \texttt{Opus-4.8} & \textbf{46.00\%} & 10.21 & 195.90 \\
 & Claude Code \texttt{Sonnet-5} & 35.00\% & 14.91 & 181.70 \\
 & Codex \texttt{gpt-5.6-sol} & 33.00\% & 4.41 & 55.61 \\
 & Codex \texttt{gpt-5.6-terra} & 42.00\% & 2.80 & 59.79 \\
\midrule
\multirow{5}{*}{\swebench{} Multilingual} & Base model & 7.00\% & -- & -- \\
 & Claude Code \texttt{Opus-4.8} & 5.00\% & 12.68 & 195.70 \\
 & Claude Code \texttt{Sonnet-5} & \textbf{22.00\%} & 14.20 & 363.77 \\
 & Codex \texttt{gpt-5.6-sol} & 15.00\% & 5.99 & 78.64 \\
 & Codex \texttt{gpt-5.6-terra} & 6.00\% & 5.18 & 56.87 \\
\midrule
\multirow{5}{*}{\swebench{} Pro} & Base model & 0.00\% & -- & -- \\
 & Claude Code \texttt{Opus-4.8} & 2.00\% & 2.19 & 17.54 \\
 & Claude Code \texttt{Sonnet-5} & 4.00\% & 9.54 & 45.63 \\
 & Codex \texttt{gpt-5.6-sol} & \textbf{9.00\%} & 11.61 & 300.49 \\
 & Codex \texttt{gpt-5.6-terra} & 1.00\% & 3.85 & 34.39 \\
\midrule
\multirow{5}{*}{GPQA Diamond} & Base model & 61.00\% & -- & -- \\
 & Claude Code \texttt{Opus-4.8} & 56.00\% & 5.98 & 27.92 \\
 & Claude Code \texttt{Sonnet-5} & 52.00\% & 6.43 & 16.03 \\
 & Codex \texttt{gpt-5.6-sol} & \textbf{65.00\%} & 2.42 & 10.37 \\
 & Codex \texttt{gpt-5.6-terra} & 64.00\% & 1.14 & 4.80 \\
\midrule
\multirow{5}{*}{AIME 2026} & Base model & 30.00\% & -- & -- \\
 & Claude Code \texttt{Opus-4.8} & 40.83\% & 4.89 & 61.36 \\
 & Claude Code \texttt{Sonnet-5} & 49.17\% & 9.29 & 121.21 \\
 & Codex \texttt{gpt-5.6-sol} & \textbf{53.33\%} & 8.90 & 63.65 \\
 & Codex \texttt{gpt-5.6-terra} & 33.33\% & 1.85 & 8.50 \\
\midrule
\multirow{5}{*}{Terminal-Bench~2.0} & Base model & 1.12\% & -- & -- \\
 & Claude Code \texttt{Opus-4.8} & 10.11\% & 8.99 & 175.63 \\
 & Claude Code \texttt{Sonnet-5} & 5.62\% & 8.87 & 156.93 \\
 & Codex \texttt{gpt-5.6-sol} & \textbf{20.22\%} & 9.56 & 69.07 \\
 & Codex \texttt{gpt-5.6-terra} & 12.36\% & 6.67 & 186.96 \\
\bottomrule
\end{tabular}%
}
\end{table}

\paragraph{Resource analysis.}
The resource columns show that high official scores are not obtained at a uniform cost. The 24 main runs have a median total wall-clock time of 6.55 hours and a median valid-candidate \tinker{} cost of \$62.51, but the spread is large: time ranges from 1.14 to 14.91 hours, and valid-candidate cost ranges from \$4.80 to \$363.77. Several high-scoring runs are comparatively inexpensive. Codex \texttt{gpt-5.6-sol} reaches 65.00\% on GPQA~Diamond in 2.42 hours with \$10.37 of \tinker{} cost, and Codex \texttt{gpt-5.6-terra} reaches 64.00\% on the same benchmark in 1.14 hours with \$4.80. Other settings consume far more budget for modest or weak outcomes: Claude Code \texttt{Sonnet-5} on Terminal-Bench~2.0 uses 8.87 hours and \$156.93 but scores 5.62\%, while Codex \texttt{gpt-5.6-sol} on \swebench{} Pro spends \$300.49 to reach 9.00\%. These contrasts show that official score should be interpreted together with search time and training cost.

\subsection{Iterative evolution and candidate selection}
\label{sec:iteration_results}
\begin{figure}[t]
\centering
\includegraphics[width=\linewidth]{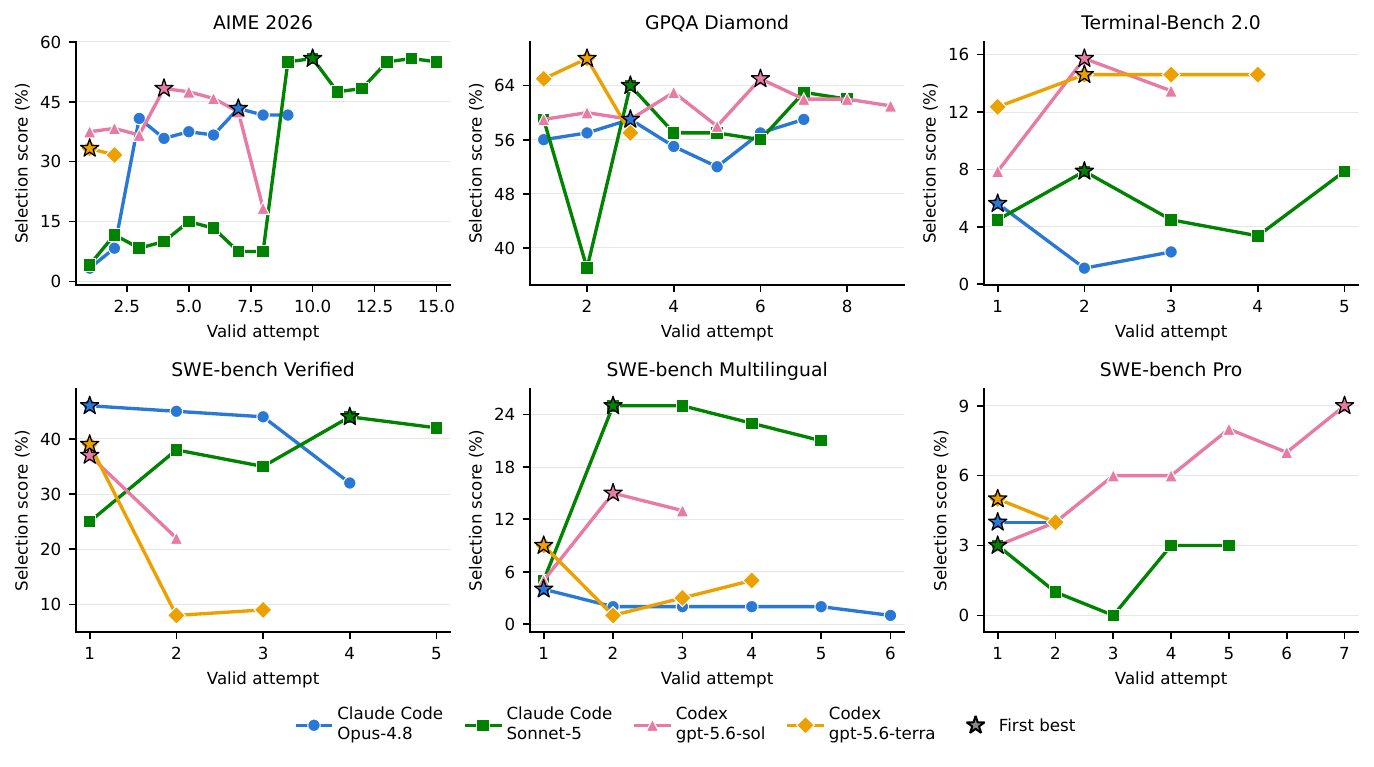}
\caption{Selection-score trajectories across valid attempts in their original order. Stars mark the first candidate that reaches the historical-best selection score.}
\label{fig:trajectories}
\end{figure}

\paragraph{Iteration improves the candidate frontier in most settings.}
A final official score evaluates only the selected checkpoint and therefore does not reveal whether iterative feedback improved upon the agent's first valid candidate. We therefore compare the first and best valid candidates across the 24 settings with valid official evaluations, excluding failed attempts and attempts without selection scores. We evaluate all candidates within a run using the same selection protocol. A later candidate outperforms the first in 14 of 24 settings, whereas in the other 10 settings the first candidate remains the best. Thus, iterative feedback often helps agents discover a stronger candidate, but not in every setting (Figure~\ref{fig:trajectories}). Official evaluation is conducted separately on the selected checkpoint.\vspace{-8pt}

\paragraph{Feedback-driven improvement is rarely monotonic.}
Among the 23 settings that continue after first reaching their best selection score, 18 finish with a final attempted candidate below that peak and the remaining 5 only return to it. This statistic describes the sequence of attempted candidates, not the checkpoint ultimately submitted: historical-best selection can preserve an earlier peak even when later attempts score lower. The result instead shows that agents do not consistently turn additional feedback into better training-data strategies. In some settings the first candidate remains strongest; in others, an intermediate improvement is followed by weaker revisions. Historical-best selection mitigates the effect of such regressions on final submission, while effective stopping and strategy revision remain necessary for efficient search.\vspace{-8pt}

\paragraph{The value and speed of iteration are strongly benchmark dependent.}
To distinguish where iteration helps from how quickly it helps, we compare first-to-best gains and time to peak. AIME~2026 improves for three of four agents, with a median gain of 25.42 percentage points: Claude Code \texttt{Sonnet-5} stays below 15.00\% for eight valid attempts before jumping to 55.00\% and peaking at 55.83\%, whereas Codex \texttt{gpt-5.6-terra}'s first candidate is already its best. GPQA~Diamond begins from a stronger frontier (56.00\%--65.00\%) and gains only 4.00 points at the median, even though all four agents refresh their first-candidate score. Iteration value therefore reflects both available headroom and how efficiently an agent converts feedback into a new training-data strategy.\vspace{-8pt}

\paragraph{Broad task families do not determine a common evolution trajectory.}
To test whether task families induce common search dynamics, we compare the three SWE-style profiles and all four agents on Terminal-Bench~2.0. \swebench{} Verified improves for one agent, \swebench{} Multilingual for two, and \swebench{} Pro accumulates gains only for Codex \texttt{gpt-5.6-sol}. On Terminal-Bench~2.0, Claude Code \texttt{Sonnet-5} improves from 4.49\% to 7.87\%, Codex \texttt{gpt-5.6-sol} recovers from an initial failed attempt and rises from 7.87\% to 15.73\% before falling to 13.48\%, Codex \texttt{gpt-5.6-terra} improves from 12.36\% to a 14.61\% plateau, and Claude Code \texttt{Opus-4.8} peaks immediately at 5.62\%. These differences make the agent--benchmark interaction, rather than the family label, the relevant unit of analysis.\vspace{-8pt}

\subsection{Budget utilization and search efficiency}
\label{sec:budget_results}
\begin{figure}[t]
\centering
\includegraphics[width=\linewidth]{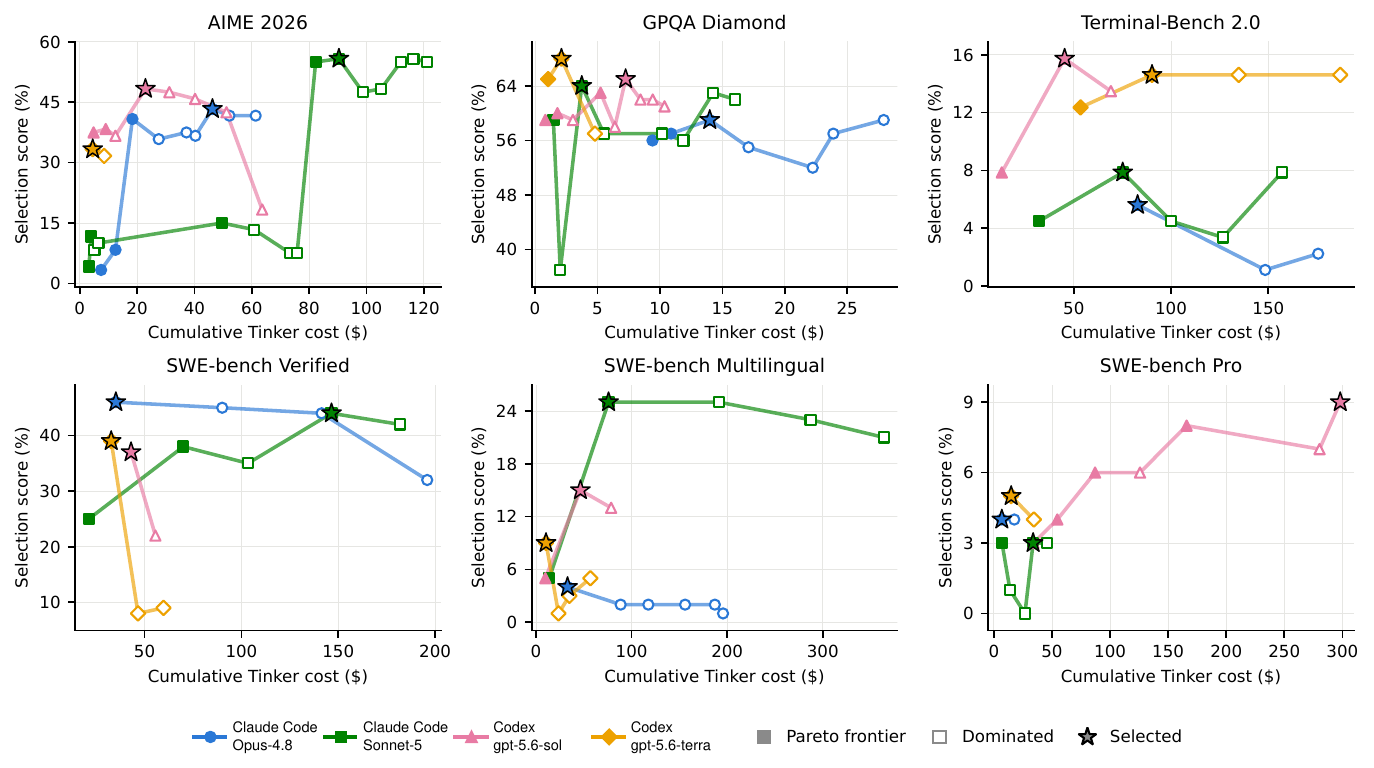}
\caption{Selection score versus cumulative \tinker{} cost across benchmarks. Filled markers indicate candidates that establish a new within-run best, hollow markers indicate the remaining candidates, and stars mark selected checkpoints.}
\label{fig:cost_pareto}
\end{figure}

\paragraph{Similar final scores can require very different search costs.}
Agents differ substantially in how much search they use. Across the 24 settings, a run has a median of 4.5 valid candidates (range: 2--15), costs \(\$62.51\) (\(\$4.80\)--\(\$363.77\)), and takes 6.55 hours (1.14--14.91). Final score alone therefore does not measure efficiency: two runs can reach similar scores even if one uses much more time and money. Figure~\ref{fig:cost_pareto} compares each candidate's selection score with the cumulative \tinker{} cost incurred up to that attempt and highlights candidates that establish a new within-run best.\vspace{-8pt}

\paragraph{Productive search can turn into costly post-peak regression.}
To test whether early frontier gains remain productive, we examine the completed Terminal-Bench~2.0 trajectories before and after their best candidate. Codex \texttt{gpt-5.6-sol} first recovers from a failed attempt, improves from 7.87\% to 15.73\% by \$45.28 cumulative cost, and then falls to 13.48\% after reaching \$69.07; its selected checkpoint nonetheless reaches 20.22\% officially with the lowest valid-candidate \tinker{} cost among the four Terminal-Bench runs. Claude Code \texttt{Sonnet-5} improves from 4.49\% to 7.87\% by \$75.16 but spends another \$81.77 without refreshing the frontier and obtains a 5.62\% official score, while Codex \texttt{gpt-5.6-terra} reaches a 14.61\% selection plateau by \$90.23 and then spends another \$96.74 for a 12.36\% official score. Search can therefore discover better candidates, yet continued spending after a peak may fail to refresh the frontier, and higher cost need not imply a stronger final checkpoint.\vspace{-8pt}

\paragraph{Checkpoint selection is itself an efficiency decision.}
To separate efficient discovery from efficient deployment, we compare the selected checkpoint with the historical-best selection score. Across the 12 GPQA, AIME, and Terminal-Bench settings, every run selects a checkpoint that attains its best observed selection score, so the development-stage selection rule does not miss the measured frontier. Selection peaks nevertheless differ from official evaluation scores by 3.89 percentage points on average: Claude Code \texttt{Sonnet-5} on GPQA falls from a 64.00\% selection peak to 52.00\% officially, while Codex \texttt{gpt-5.6-sol} on AIME rises from 48.33\% to 53.33\% and Codex \texttt{gpt-5.6-sol} on Terminal-Bench rises from 15.73\% to 20.22\%. The trajectories further show that selection discipline can save substantial post-peak spending: the Terminal-Bench runs spend \$23.79--\$96.74 after first reaching their best selection score without improving it. These cases show that stopping and checkpoint choice determine whether search expenditure is converted into the best discovered model.\vspace{-8pt}

\subsection{Mechanism analysis}
\label{sec:research_mechanisms}
\paragraph{Mechanism analysis links score changes to concrete data-synthesis research decisions.}
To move beyond final-score comparisons, we code each attempt by its diagnosed capability gap, data source, validation signal, and supervision while the training and evaluation stack remains fixed. We compare these decisions with selection and official scores to identify the cross-case mechanisms in Table~\ref{tab:evolution_mechanisms}. Appendix~\ref{app:evolution_cases} provides the full trajectories and case-level analysis.

\begin{table}[t]
\caption{Cross-case mechanisms suggested by the current process evidence.  These are process-level observations, not single-factor causal ablations.}
\label{tab:evolution_mechanisms}
\centering
\footnotesize
\begin{tabularx}{\linewidth}{lXX}
\toprule
Mechanism & Observation & Implication for researcher agents \\
\midrule
Diagnose the real capability gap & Claude Code \texttt{Sonnet-5} on AIME stays below 15.00\% for eight valid attempts before jumping to 55.00\% and then 55.83\%. & Agents should recognize when local revisions are not changing the effective data strategy and search for a qualitatively different capability hypothesis. \\
Embed validation signals & AIME uses answer-agreement filtering, \swebench{} Pro uses environment-successful trajectories, and Terminal-Bench~2.0 uses sandbox execution with explicit task outcomes. & Agents should use executable or decidable checks rather than relying only on model self-judgment. \\
Match supervision to target behavior & Real submit behavior improves \swebench{} Pro; on Verified, final-action data reaches a 39.00\% selection score while all-turn and safe variants reach 8.00\% and 9.00\%; Terminal-Bench~2.0 improves for Codex \texttt{gpt-5.6-sol} after a more focused candidate but regresses after a later stop-oriented modification. & Supervision granularity and action source are data-design variables, although training changes prevent a single-factor claim. \\
Preserve historical best candidates & Claude Code \texttt{Sonnet-5} on AIME reaches its first 55.83\% peak at the tenth valid candidate and later returns to but does not exceed it; Codex \texttt{gpt-5.6-sol} on \swebench{} Pro peaks at the ninth attempted round; Codex \texttt{gpt-5.6-sol} on Terminal-Bench peaks at its second valid candidate before regressing. & Agents need both continued search and explicit stop or rollback policies. \\
Reasoning effort changes search depth and resource allocation & On \swebench{} Verified with Claude Code \texttt{Sonnet-5}, max effort improves the first candidate from 25\% to 36\%, the best selection score from 44\% to 49\%, and the official score from 35\% to 52\%. & Effort should be analyzed as a research control that changes the search policy, not only as extra agent-side computation. \\
\bottomrule
\end{tabularx}
\end{table}

\begin{wrapfigure}{r}{0.4\linewidth}
\vspace{-10pt}
\centering
\includegraphics[width=\linewidth]{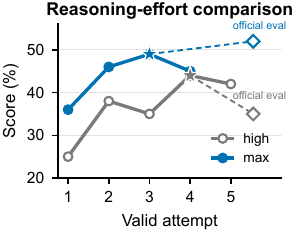}
\caption{Reasoning-effort diagnostic on \swebench{} Verified with the Claude Code \texttt{Sonnet-5} researcher agent. The figure compares within-run selection scores under high and max effort. Stars mark the first historical-best candidate, and diamonds show the selected checkpoint's official evaluation score.}
\label{fig:reasoning_effort}
\vspace{-10pt}
\end{wrapfigure}

\paragraph{Higher reasoning effort improves this diagnostic setting by changing the training-data research policy.}
We compare the Claude Code \texttt{Sonnet-5} researcher agent on \swebench{} Verified under high and max reasoning effort, holding the rollout model, target model, benchmark subset, budgets, and training--evaluation pipeline fixed.  Under this diagnostic (Figure~\ref{fig:reasoning_effort}), max effort raises the mean selection score over the first four comparable attempts from 35.5\% to 44.0\%, improves the first candidate from 25\% to 36\%, and moves the historical best from 44\% to 49\%.  The official score also increases from 35\% to 52\%, although this single-run difference should not be interpreted as a fully isolated causal estimate.

The resource profile clarifies the mechanism.  Compared with high effort, max effort completes one fewer valid attempt, but its selected dataset is substantially larger: 149 training records rather than 50, 2,681 trainable assistant turns rather than 595, and \(\$401.70\) rather than \(\$181.70\) in candidate-loop \tinker{} cost.  The result is therefore consistent with a depth--breadth tradeoff: higher effort finds a stronger candidate earlier and constructs a more resource-intensive training-data program, while reducing the number of candidates that fit within the fixed budget.

\paragraph{Effective evolution combines sound data decisions with disciplined search.}
Across the main trajectories, improvements are associated with diagnosing the target capability, grounding data in validation signals, matching supervision to evaluated behavior, and preserving strong checkpoints. The reasoning-effort diagnostic further suggests that deeper data construction can trade search breadth for candidate quality. Together, these patterns support closed-loop training-data research rather than imitation-based generation.

\subsection{Early Experiment of RSI}
\label{sec:early_rsi_k26}

\begin{figure}[H]
\centering
\includegraphics[width=\linewidth]{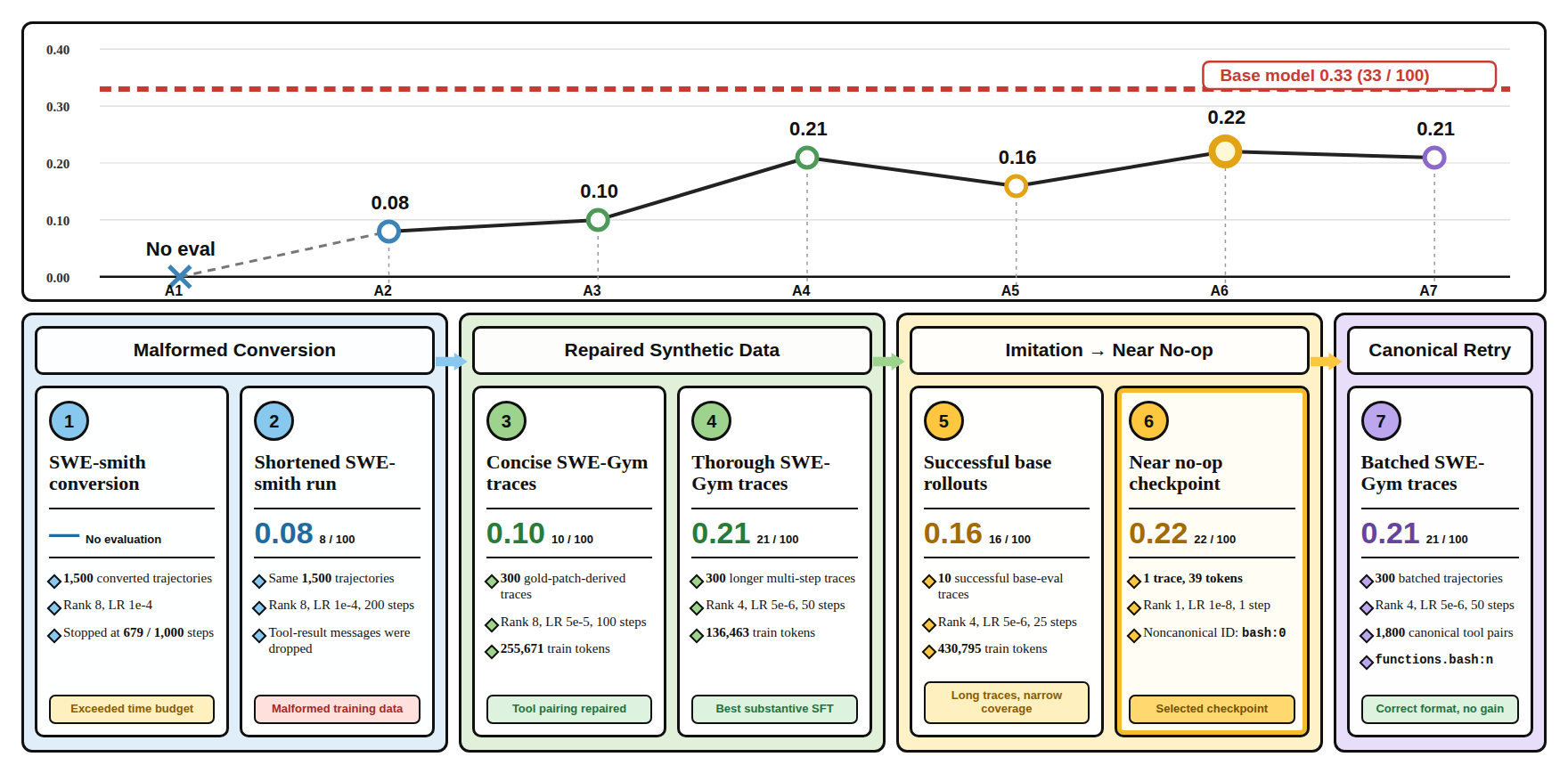}
\caption{Early RSI trajectory on \swebench{} Pro. The Kimi K2.6 researcher evolves its synthetic-data strategy across seven \tinker{} LoRA attempts, improving the candidate score but remaining below the 33\% unadapted-model reference.}
\label{fig:early_rsi_k26}
\end{figure}

\paragraph{Setting.}
We run an exploratory same-family RSI experiment outside the main matrix. A Claude Code harness powered by \texttt{kimi-k2.6} serves as the researcher agent, the target is the instruction-tuned \texttt{moonshotai/Kimi-K2.6} model on \tinker{}, and Claude Opus~4.8 is the fixed rollout model. Candidates are trained with LoRA SFT and evaluated by \miniswe{} through \harbor{}--\eibtwo{} on a fixed 100-task \swebench{} Pro subset, with concurrency 64 and a 200-step limit. The backend uses a 32,768-token context window, a 28,671-token prompt limit, and up to 4,096 output tokens. The run has a 20-hour and \$2,000 budget and uses 12.3 hours and an estimated \$738.69. Figure~\ref{fig:early_rsi_k26} reports within-run candidate evaluations.

\paragraph{Results.}
The trajectory shows clear strategy evolution but no effective model improvement. The agent first converts 1,500 resolved SWE-smith trajectories into full assistant--tool conversations and schedules 1,000 high-learning-rate steps, but stops at step 679 without evaluation. It then shortens the same recipe to 200 steps, scoring 8\%; inspection shows that the converter dropped tool-result messages and produced malformed supervision. The next two attempts switch to 300 gold-patch-derived SWE-Gym trajectories with repaired tool pairing: concise traces score 10\%, while longer multi-step traces combined with a lower learning rate and LoRA rank improve to 21\%. The agent then extracts 10 successful unadapted-model rollouts as imitation data, but their narrow coverage reduces the score to 16\%. To test whether training itself is destructive, it creates a near-no-op adapter from one 39-token example, one step, rank 1, and a negligible learning rate; this reaches 22\% and is selected. A final attempt returns to 300 SWE-Gym trajectories, batches commands into 1,800 canonical tool pairs, and changes identifiers from forms such as \texttt{bash:0} to \texttt{functions.bash:n}, but scores 21\%. 

In conclusion, Although agent has improved the synthetic data pipeline and improve the performance from 8\% to 21\%, every checkpoint remains below the 33\% unadapted reference.  This likely because the target is already a capable instruction model on an agentic benchmark and the current researcher's RSI capability remains insufficient to discover an improving training distribution.

%% file: sections/discussion.tex
\section{Discussion and Limitations}
\label{sec:discussion}

\paragraph{Official evaluation does not yet establish adaptive generalization.}
Our evaluation prioritizes broad coverage under substantial sandbox and training costs, using one representative run per setting and fixed subsets for four benchmarks. Checkpoint selection and official evaluation use the same task subset, so the reported scores measure performance in fresh environments but not generalization to statistically held-out tasks. Repeated trials, evaluator-private splits, and refreshed tasks are needed to measure the stability and transferability of agentic training-data research strategies. Researcher identity also bundles the underlying LLM, agent scaffold, and configured reasoning effort, so the main matrix compares complete researcher systems rather than isolating any one component.

% \paragraph{The benchmark isolates one component of recursive self-improvement.}
% \benchmark{} evaluates a bounded, cross-model setting in which a researcher agent improves a separate target model through iterative data-centric post-training. It does not evaluate self-modification of the researcher agent or open-ended recursive improvement.

% \textcolor{red}{\textbf{TODO:} Strengthen the evidence with a small equal-budget feedback-versus-no-feedback study and repeated runs on representative agent--benchmark settings.}

%% file: sections/conclusion.tex
\section{Conclusion}
\label{sec:conclusion}

We introduced \benchmark{}, a controlled benchmark for evaluating the data-centric research capability required for recursive self-improvement under a fixed post-training stack. Across four frontier agents and six benchmarks, agents exhibit some core capabilities of data-centric researchers: iteration improves on the first valid attempt in 58.33\% of settings. Yet agents do not improve consistently from feedback: among searches that continue after reaching their best observed score, 78.26\% finish with a lower-scoring final attempt and the rest only recover the same peak. Historical-best selection can protect the submitted checkpoint from these regressions, but it does not make the underlying research process reliably progressive. Together, these results reveal a discovery--reliability gap in how agents translate feedback into better training-data strategies. We hope \benchmark{} can help develop more dependable data-centric researcher agents for recursive self-improvement by making their research process measurable and auditable.

%% file: sections/appendix_action_space.tex
\newpage
\begin{center}
  \LARGE\textbf{Appendix}
\end{center}

\section{Evaluation Suite and Seed Repositories}
\label{app:evaluation_suite}

The complete evaluation protocol used in the main experiments is listed in Table~\ref{tab:evaluation_suite}. All agents within a benchmark use the same evaluation split, sampling protocol, runner, verifier, and decoding settings. The final column lists the public seed repositories or data sources available to the researcher agents for constructing training experience.

\begin{table}[H]
\caption{Evaluation suite and seed repositories. The metric matches each benchmark's success criterion, and the same evaluation split and sampling protocol are used for every agent. AIME uses four decoding runs per problem.}
\label{tab:evaluation_suite}
\centering
\footnotesize
\resizebox{\linewidth}{!}{%
\begin{tabular}{llllllp{3.2cm}}
\toprule
Benchmark & Evaluation items & Original size & Coverage & Evaluation runner & Metric & Seed repositories \\
\midrule
\swebench{} Verified & 100 & 500 & Fixed subset, 20.0\% & \miniswe{} & Resolved rate & SWE-smith; SWE-Gym; swe-factory \\
\swebench{} Multilingual & 100 & 300 & Fixed subset, 33.3\% & \miniswe{} & Resolved rate & SWE-smith; SWE-Gym; swe-factory \\
\swebench{} Pro & 100 & 731 & Fixed subset, 13.7\% & \miniswe{} & Resolved rate & SWE-smith; SWE-Gym; swe-factory \\
Terminal-Bench 2.0 & 89 & 89 & Full set & \textsc{Terminus-2} & Task success rate & endless-terminals; tmax \\
GPQA Diamond & 100 & 198 & Fixed subset, 50.5\% & \textsc{Terminus-2} & Accuracy & synthetic-data-kit \\
AIME 2026 & 30 ($4\times$ each) & 30 & Full set & \textsc{Terminus-2} & Accuracy & synthetic-data-kit \\
\bottomrule
\end{tabular}%
}
\end{table}

\section{Representative Actions Within the Data-Synthesis Research Loop}
\label{app:action_space}

The operations below elaborate the agent-controlled stages of the loop in Figure~\ref{fig:framework}(c). They define a representative action space, not a required modular architecture: an agent may implement any subset, combine several operations in one procedure, or introduce alternatives that respect the benchmark's input, integrity, training, and evaluation contracts.

\subsection{Diagnose the Next Training-Data Need}

The agent can organize available failures by capability, error type, task family, environment, tool, interaction length, or stage of execution. It may distinguish missing knowledge from failures of reasoning, planning, tool use, state tracking, recovery, verification, or task completion, and use base-model or prior-attempt behavior to estimate where additional supervision is most likely to transfer. The resulting diagnosis specifies a concrete target for the next dataset rather than treating data volume as the solution.

Diagnosis can also compare the intended lesson of a previous dataset with the behavior actually changed by training. A failed attempt may indicate that the original capability hypothesis was wrong, that the generated examples did not express the intended behavior, that invalid examples survived filtering, or that the supervision format did not match the target evaluation protocol. These alternatives imply different revisions even when they produce the same aggregate score.

\subsection{Synthesize Targeted Training Experience}

\paragraph{Tasks and executable states.}
The agent may generate synthetic tasks that exercise capabilities implicated by its diagnosis. For interactive tasks, it may also construct the initial state required to execute them, including files and repositories, databases, terminal or filesystem states, tool configurations, or other environment state. Generated tasks should satisfy the target interaction contract and expose behavior that can be exercised by the fixed evaluation service.

Task construction can vary capability coverage and composition rather than only surface wording. An agent may target particular error types, repositories, tools, or interaction lengths; create controlled variants of existing task families; or combine permitted public sources with newly synthesized instances. These choices determine which regions of the target behavior receive supervision.

\paragraph{Successful, failed, and recovery trajectories.}
For long-horizon tasks, the agent may generate trajectories containing actions, observations, intermediate states, and a completion signal. Successful or expert trajectories demonstrate behavior that reaches a verified outcome. Failed trajectories expose invalid tool calls, incorrect plans, state-tracking errors, premature completion, or other recurrent failure modes. Recovery and repair trajectories demonstrate how to recognize a failed state, revise the plan, and complete the task after receiving new evidence.

The supervision need not reproduce every interaction turn. An agent may train on complete trajectories, selected decision points, final actions, critiques, repaired continuations, or mixtures of these representations. It may also decide which messages contribute to the training loss and which remain as context. The relevant design choice is whether the representation teaches behavior required by the target evaluation protocol without overwhelming training with irrelevant interaction history.

\subsection{Validate and Organize Training Experience}

\paragraph{Verification, filtering, and deduplication.}
Candidate tasks and trajectories may be checked through executable environments, agent-designed or permitted auxiliary verifiers, answer agreement, schema validation, consistency checks, or other decidable signals. The agent may discard examples that are invalid, unverifiable, contaminated, malformed, or inconsistent with their claimed outcome, and may deduplicate exact, near-duplicate, or structurally redundant examples. An auxiliary verifier or checker is therefore useful not as an isolated component, but as a mechanism for deciding which proposed experiences are trustworthy enough to influence the model.

\paragraph{Long-trajectory representation.}
Agentic trajectories often contain long observations, tool outputs, test logs, and repeated interaction history. The agent may truncate, compress, summarize, segment, or reconstruct these contexts while preserving the state and evidence needed for the supervised decision. Representation choices include the granularity of supervised actions, the placement of verifier observations, the treatment of failed turns, and the protocol used to express task completion.

\paragraph{Difficulty, curriculum, and mixture.}
The agent may estimate difficulty using task metadata, execution length, tool-use complexity, verifier outcomes, generator disagreement, base-model performance, or other available signals. It can use these estimates to balance easy, medium, and hard examples, order their training exposure, or focus on the region where the model is most likely to improve. Mixture decisions determine the proportions of task families, capability categories, trajectory types, supervision formats, and difficulty levels, including the balance among successful behavior, failure recognition, recovery, verification, and task completion.

\subsection{Revise the Strategy and Manage Search}

After each shared training and evaluation cycle, the agent may revise any preceding decision. It can replace a data source, narrow or expand task coverage, change the trajectory representation, strengthen an auxiliary data checker, alter filtering thresholds, adjust training exposure, or rebalance the curriculum and mixture. A useful revision follows from a diagnosis of the previous checkpoint rather than from an assumption that another attempt or a larger dataset will necessarily help.

These revisions compete for the remaining wall-clock and training-cost budgets. The agent must decide whether to exploit a promising recipe, explore a different hypothesis, return to an earlier strategy, preserve a historical-best checkpoint, or stop. Budget allocation, rollback, stopping, and final checkpoint selection are consequently part of the same iterative policy.

\subsection{Instantiation Across Benchmark Families}

For software-engineering benchmarks, the loop may connect repository state, issue interpretation, code localization, patch generation, test execution, review, verification, and submission behavior. For terminal tasks, it may connect initial filesystem or tool state, command sequences, state changes, recovery from unsuccessful commands, verification observations, and completion timing. For science question answering and mathematics, the emphasis may shift toward problem construction, reasoning traces, answer verification, generator agreement, difficulty control, and the selection of which reasoning steps receive supervision.

These examples illustrate why one generic synthetic-data recipe is unlikely to transfer uniformly across the suite. The appropriate training-data strategy is conditioned on the target benchmark and current model failures, while the surrounding SFT and evaluation machinery remains fixed.

\subsection{Representative Evolution Cases}
\label{app:evolution_cases}

\begin{figure}[t]
\centering
\includegraphics[width=\linewidth]{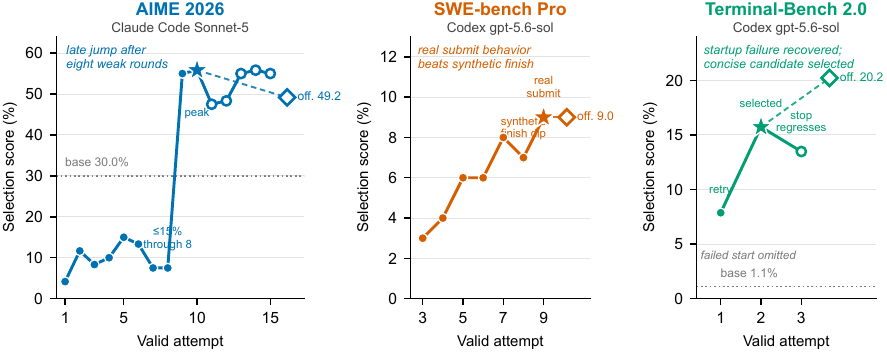}
\caption{Representative evolution trajectories on AIME~2026, \swebench{} Pro, and Terminal-Bench~2.0. Circles show valid-attempt selection scores, stars mark selected historical-best candidates, and open diamonds show official evaluations.}
\label{fig:evolution_cases}
\end{figure}

We select three runs with valid official evaluations, at least three valid candidates, and auditable data changes. We plot their quoted selection and official scores in Figure~\ref{fig:evolution_cases} and trace the strategy changes associated with each trajectory below.

\paragraph{Late strategy shifts can produce discrete AIME gains.}
Claude Code \texttt{Sonnet-5}'s AIME trajectory shows that improvement need not be gradual. Its first eight valid candidates remain weak, never exceeding a 15.00\% selection score. The ninth candidate jumps to 55.00\%, and the tenth reaches 55.83\%; later candidates fluctuate between 47.50\% and 55.83\% without surpassing that peak. The selected tenth checkpoint obtains 49.17\% officially, above the 30.00\% base-model reference but below the development-stage peak. This trajectory suggests a stage transition in the data factory: repeated local revisions before the jump were insufficient, while the later strategy exposed a much stronger reasoning-training regime. The score sequence alone does not isolate which component caused the jump, so we treat it as process evidence for qualitative strategy change rather than a single-factor ablation.

\paragraph{Behavior-aligned trajectories outperform synthetic completion shortcuts in the \swebench{} Pro case.}
To test whether environment-successful repairs provide sufficient supervision, we compare how Codex \texttt{gpt-5.6-sol} represents task completion on \swebench{} Pro. The agent begins with repair trajectories that succeed in execution, then uses evaluation failures to build a curriculum that teaches the model to review its final diff, verify its edits, and submit the task. Its valid selection scores progress from 3.00\% to 4.00\%, 6.00\%, 6.00\%, and 8.00\%; a variant supervised on synthetic finish actions regresses to 7.00\%, whereas grounding the closing behavior in real submissions from successful trajectories reaches 9.00\% and is selected for official evaluation. This contrast associates stronger SWE performance with supervision that represents the complete tool-use and task-closing behavior rather than only the existence of a correct patch.

\paragraph{Failure recovery and focused supervision separate validity from performance in Terminal-Bench~2.0.}
Codex \texttt{gpt-5.6-sol}'s Terminal-Bench~2.0 trajectory first produces a candidate that fails before yielding a usable selection score. After recovery, the first valid candidate reaches 7.87\%. A more focused second valid candidate raises the selection score to 15.73\% and is selected, while a later stop-oriented modification regresses to 13.48\%. The selected checkpoint obtains 20.22\% officially (18/89), compared with the 1.12\% base-model reference. This trajectory separates three abilities: recovering a failed training run creates a valid candidate, focused supervision advances the frontier, and historical-best preservation prevents a later targeted modification from overwriting the stronger checkpoint.